\documentstyle[aps,prb,preprint]{revtex} 

\begin{document}  \preprint{\today} \draft
\newcommand{\bk}{{\bf k}}
\newcommand{\bq}{{\bf q}}
\newcommand{\br}{{\bf r}}
\title{Static and Dynamic Spin Structure Factors of
Frustrated Heisenberg Antiferromagnet}

\author{O. P. Sushkov\cite{Budker} }

\address{ School of Physics, The University of New South Wales,
    Sydney 2052, Australia}
\maketitle

\begin{abstract}
Using the modified spin-wave theory we calculate static and dynamic
spin structure factors in spin-liquid state of the $J_1$-$J_2$ model.
The spectrum of excitations in the vector channel is discussed.
The developed technique can also be applied to the $t-J$
model describing copper oxide superconductors.
\end{abstract}

\narrowtext 
Magnetically disordered states in quantum spin models
are of  considerable interest.
Much of this interest stems from the connection of this problem to
 high-$T_c$ superconductivity.
The ground state of an undoped compound has long range antiferromagnetic
order, which is well described by the Heisenberg model and has been studied by
 numerous methods.\cite{Manousakis}
However introducing a small number of holes leads to destruction of
 long range order.

Destruction of long range order can be studied by introducing some
frustration into the Heisenberg model. This simplified
mechanism of the destruction  is certainly far from the
reality for cuprates, but the investigation
of this relatively simple model allows us to develop approaches
to more realistic models. From this point of view
the main purpose of the present work is to develop a technique
applicable to the $t-J$ model.

We will focus on the simplest frustrated Heisenberg model.
It is the $J_1$-$J_2$ model defined by
\begin{equation} 
\label{H}
  H = J_1 \sum_{\rm NN} {\bf S}_\br \cdot {\bf S}_{\br'}
    + J_2 \sum_{\rm NNN} {\bf S}_\br \cdot {\bf S}_{\br'}.
\end{equation}
In this Hamiltonian, the $J_1$ term describes 
 the usual Heisenberg interaction of the nearest neighbor spins 
($S={1\over2}$) on a square lattice,
 while the $J_2$ term introduces a frustrating interaction
  between the next nearest neighbor sites.
For convenience, we set $J_1 = 1$ and denote $\alpha \equiv J_2 / J_1$

For small $\alpha$, the ground state is N\'{e}el ordered. For large $\alpha$
the system is decomposed into two N\'{e}el ordered sublattices which,
however, have the same quantization axis. This is so called
collinear state. Whether or not the N\'{e}el and collinear states are
separated in parameter space by a spin liquid state has been a
subject of many discussions. Besides the spin-wave 
calculations\cite{linear,mean_field,spin-waves,Igarashi,Gochev},
the model has been studied by the Schwinger boson mean field 
theory\cite{Schwinger},
analysis of small lattices\cite{numerical},
a series expansion\cite{series},
a mean field theory of bond operators\cite{SB}, and other 
methods\cite{other}.
The majority of these works including most recent exact finite lattice 
dioganalizations\cite{Sch94} and series expansion\cite{Oit} indicate 
the existence of a different intermediate phase for $0.4 < \alpha < 0.6$. 
Therefore in this work we assume that the system undergoes a second order 
quantum phase transition at $\alpha = \alpha_c \approx 0.4$ from the N\'{e}el 
to a spin-liquid state.
We are interested in this spin liquid phase close to the transition point,
so $\alpha > \alpha_c$ and $\alpha-\alpha_c \ll 1$. 
In our previous work\cite{Dot} for description of this state 
we used modified spin-wave theory originally suggested by 
Takahashi\cite{Takahashi} for the Heisenberg model at nonzero temperature.
In the present work we use same approach.

We remind the reader that for spin-wave description it is convenient
to use Dyson-Maleev transformation\cite{DM} (see also 
review\cite{Manousakis}) for localized spin $S=1/2$,
\begin{eqnarray}
\label{dm}
&&S_l^-=a_l^{\dag}, \ \ \ S_l^+=(2S-a_l^{\dag}a_l)a_l,\nonumber\\
&&S_l^z=S-a_l^{\dag}a_l, \ \ \ for\ \ l \in up \ \ \ sublattice;\\
&&S_m^-=b_m, \ \ \ S_m^+=b_m^{\dag}(2S-b_m^{\dag}b_m),\nonumber\\
&&S_m^z=-S+b_m^{\dag}b_m, \ \ \ for\ \ m \in down \ \ \ sublattice,
\nonumber
\end{eqnarray}
and Fourier representation for $a_l$ and $b_m$
\begin{eqnarray}
\label{Fu}
a_l&=&\sqrt{2\over{N}}\sum_{\bk}e^{i{\bk}{\br}_l}a_{\bk}\\
a_m&=&\sqrt{2\over{N}}\sum_{\bk}e^{i{\bk}{\br}_m}b_{\bk}.\nonumber
\end{eqnarray}
Here $N$ is number of sites on square lattice. The summation over
${\bk}$, here and everywhere below is restricted to the inside magnetic 
Brillouine zone ($|k_x|+|k_y| \le \pi$). 
There are two simple ways to find an effective Hamiltonian quadratic 
in the operators $a$ and $b$ . First way is just dropping the
quartic terms, and this is the linear spin-wave theory (LSWT).
Second way corresponds to the mean field treating of the quartic
terms 
$a^{\dag}a b^{\dag}b \to
\langle a^{\dag}a\rangle b^{\dag}b +
\langle a^{\dag}b^{\dag}\rangle  ab + ...$, and this is mean field
spin-wave theory (MFSWT). Both approximations give very similar
results (see e. g. Ref.\cite{Dot}), Therefore as soon as we believe that
spin-liquid phase exists, it does not matter which
approximation is used for calculations of the properties of this phase.
However LSWT gives the value of $\alpha_c \approx 0.4$ very close to that
found from exact numerical computations. Therefore in the present work
we will use linear spin-wave theory. 

Using  Bogoliubov 
transformation one can easily find spectrum of spin-waves and staggered
magnetization of sublattice in the N\'{e}el state
\begin{eqnarray}
\label{on}
\omega_{\bk}&=&2\sqrt{[1-\alpha (1-\eta_{\bk})]^2-\gamma_{\bk}^2},\\
m&\equiv&|\langle S_z \rangle| = 1-{2\over N}\sum_{\bk}
{{1-\alpha (1-\eta_{\bk})}\over{\omega_{\bk}}}.\nonumber
\end{eqnarray}
As usually we have defined 
\[ 
  \gamma_\bk = {1\over2} (\cos k_x + \cos k_y)
 \ \  {\rm and} \ \
 \eta_\bk = \cos k_x \cos k_y.
\]

At $\alpha=0$ the staggered magnetization $m=0.3$.
It decreases with increasing of $\alpha$ and vanishes at 
$\alpha=\alpha_c \approx 0.4$ where the system undergoes a second
order transition into the liquid state.
Zero sublattice magnetization means that
 the ground state is a condensate of many spin waves
 $a_{\bf k}$ and $b_{\bf k}$.
To describe the emerging phase we must take into account their
 non-linear interaction. We cannot do this exactly.
However, there is an approximate method\cite{Dot} originally suggested by 
Takahashi\cite{Takahashi} for the Heisenberg model at nonzero temperature.
Following Takahashi, we impose an additional condition that
 sublattice magnetization is zero
\begin{equation} 
\label{constraint}
\langle S^z_u - S^z_d \rangle =
 \langle 
 \case{1}{2} - a^{\dag}_{\bf l} a_{\bf l}  +
 \case{1}{2} - b^{\dag}_{\bf m} b_{\bf m} \rangle
 = 0,
\end{equation}
where $u$ and $d$ are the spin up and down sublattices.
The constraint (\ref{constraint}) gives an effective cutoff of unphysical 
states in Dyson-Maleev transformation. This question is discussed in
the paper\cite{Dot}.

The constraint (\ref{constraint}) is introduced into the Hamiltonian
 via a Lagrange multiplier ${1\over 8}\nu^2$. 
Now we must diagonalize
\begin{eqnarray} 
\label{Hnu}
 H_{\nu}&=& H_{LSWT} - {1\over 8}\nu^2 (S^z_u - S^z_d)\to\\
&\to &
2\sum_{\bk}\left(A_{\bk}(a^{\dag}_{\bk}a_{\bk}+b^{\dag}_{\bk}b_{\bk})
+\gamma_{\bk}(a_{\bk}b_{-\bk}+a^{\dag}_{\bk}b^{\dag}_{-\bk})\right),
\nonumber
\end{eqnarray}
where $A_{\bk}=1-\alpha (1-\eta_{\bk})+{1\over 8}\nu^2$.
The simple (linear) second term in (\ref{Hnu}), taken together with
 Eq.~(\ref{constraint}), takes account of non-linear interaction
 of spin waves.
Diagonalizing Eq.~(\ref{Hnu}) by Bogoliubov transformation
\begin{eqnarray}
\label{Bog}
a_{\bk}&=&U_{\bk}\alpha_{\bk}+V_{\bk}\beta^{\dag}_{-\bk},\\
b_{-\bk}&=&V_{\bk}\alpha^{\dag}_{\bk}+U_{\bk}\beta_{-\bk},\nonumber
\end{eqnarray}
we get the spectrum of excitations
\begin{equation}
\label{onu}
\omega_{\nu \bk}=2\sqrt{A_{\bk}^2-\gamma_{\bk}^2}.
\end{equation}
This spectrum has a gap \ $\nu\sqrt{1+{{\nu^2}\over{16}}} \approx \nu$,
so the meaning of Lagrange multiplier is elucidated.
Taking also into account that in thermal equilibrium
\begin{equation}
\label{nk}
n_{\bk}\equiv \langle \alpha^{\dag}_{\bk}\alpha_{\bk}\rangle
=\langle \beta^{\dag}_{\bk}\beta_{\bk} \rangle =
{1\over{\exp(\omega_{\nu \bk}/T)-1}}
\end{equation}
we get from (\ref{constraint}) the equation for $\nu$
\begin{equation}
\label{nu}
0=1-{2\over N}\sum_{\bk}
{{A_{\bk}}\over{\omega_{\nu \bk}}}(1+2n_{\bk}).
\end{equation}

The spin-wave velocity does not vanish at critical point
$c=\sqrt{2(1-2\alpha)}\approx 0.7$. Therefore for $\nu,k \ll 1$
the spectrum is of the form $\omega_{\nu \bk}=\sqrt{\nu^2+c^2 k^2}$
and  equation (\ref{nu}) can be re-written\cite{Dot} as
\begin{equation}
\label{nu1}
m+{2\over N}\sum_{\bk}\left[{1\over{ck}}-{1\over{\sqrt{\nu^2+c^2 k^2}}}
\coth\left({{\sqrt{\nu^2+c^2 k^2}}\over{2T}}\right)\right]=0,
\end{equation}
where $m < 0$ is given by Eq.(\ref{on}).
We would like to stress that that the condition $c \ne 0$
is not crucial for the validity of the method. Moreover
for $t-J$ model, which we are mainly interested in, the speed
vanishes, or even $c^2 < 0$. Nevertheless the method works\cite{SWC}.
For $J_1-J_2$ model $c > 0$, and we will use the simplification
$\omega_{\nu \bk} \to \sqrt{\nu^2+c^2 k^2}$, but from
comparison with exact numerical solution of equation (\ref{nu})
we know that it is valid only for very small $\nu$ and $T$: 
$\nu, T \ll {1\over{10}}$.
Solution of equation (\ref{nu1}) at zero temperature is straightforward
\begin{equation}
\label{nu0}
\nu_0=\pi |m|c^2 =\pi B c^2 (\alpha-\alpha_c) \approx 3.7 (\alpha-\alpha_c).
\end{equation}
We took into account that near critical point $m=B(\alpha_c-\alpha)$
with slope $B \approx 2.4$ according to LSWT and
most recent exact finite lattice dioganalizations\cite{Sch94} and series 
expansion\cite{Oit}. Index $0$ in $\nu_0$ indicates that it is a
gap at zero temperature. Equation (\ref{nu1}) can be also
written as\cite{Dot}
\begin{equation} 
\label{gap}
2T \ln \left( 2 \sinh {\nu \over 2T} \right) = -\pi m c^2=\nu_0.
\end{equation}
Similar equation has been obtained in the Ref.\cite{Sachdev}
for the non-linear $\sigma$-model in the limit
when the number of components of the order parameter
${\cal N} = \infty$. Solution of equation (\ref{gap}) at low and high
temperature looks like
\begin{eqnarray}
\label{sol}
\nu&\approx& \nu_0(1+e^{-\nu_0/T}), \ \ \ T \ll \nu_0,\\
\nu&\approx& \Theta T +{1\over{\sqrt{5}}}\nu_0, \ \ \ \ T \gg \nu_0,\nonumber
\end{eqnarray}
where $\Theta=2\ln {{1+\sqrt{5}}\over{2}} = 0.9624$. We follow the notations
of Ref.\cite{Sachdev}.

Now we can proceed to the calculation of spin structure factor in 
spin-liquid state. It is quite similar to the Takahashi's calculation
for Heisenberg model at nonzero temperature\cite{Takahashi}.
Using Dyson-Maleev representation (\ref{dm}),
constraint (\ref{constraint}) ($\langle a^{\dag}_l(t) a_l(t)\rangle=
\langle b^{\dag}_m(t) b_m(t)\rangle ={1\over{2}}$),
and mean field procedure for averaging of the quartic terms 
($\langle a^{\dag}a b^{\dag}b\rangle \to
\langle a^{\dag}a\rangle \langle b^{\dag}b\rangle +
\langle a^{\dag}b^{\dag}\rangle \langle ab \rangle$)
one finds
\begin{eqnarray}
\label{SS}
\langle {\bf S}_m(t) {\bf S}_l(0) \rangle &=&
-\langle b^{\dag}_m(t)a^{\dag}_l(0)\rangle 
\langle b_m(t) a_l(0) \rangle,\\
\langle {\bf S}_{l^{\prime}}(t) {\bf S}_l(0) \rangle &=&
\langle a^{\dag}_{l^{\prime}}(t)a_l(0)\rangle 
\langle a_{l^{\prime}}(t) a^{\dag}_l(0) \rangle,\nonumber
\end{eqnarray}
where $l,l^{\prime} \in \ up \  sublattice$, and
$m \in down \ sublattice$. Further calculation gives
\begin{eqnarray}
\label{SS1}
\langle {\bf S}_i(t) {\bf S}_j(0) \rangle &=&f^2(t,{\bf r}_{ij})
-{1\over{4}} \delta^2(t,{\bf r}_{ij}),\ \ \
i,j \in same\ \ sublattice,\\
\langle {\bf S}_i(t) {\bf S}_j(0) \rangle &=&-g^2(t,{\bf r}_{ij}),\ \ \
i,j \in different\ \ sublattices \nonumber.
\end{eqnarray}
where
\begin{eqnarray}
\label{fgexp}
f({t,\br})&=&
{2\over {N}}\sum_{\bk}e^{i{\bf kr}}{{A_{\bk}} \over{\omega_{\nu{\bk}}}}
\left[e^{-i\omega_{\nu{\bk}}t}(1+n_{\bk})+
e^{i\omega_{\nu{\bk}}t}n_{\bk}\right],\nonumber\\
\delta({t,\br})&=&
{2\over {N}}\sum_{\bk}e^{i{\bf kr}}
\left[e^{-i\omega_{\nu{\bk}}t}(1+n_{\bk})-
e^{i\omega_{\nu{\bk}}t}n_{\bk}\right],\\
g({t,\br})&=&
-{2\over {N}}\sum_{\bk}e^{i{\bf kr}}{{\gamma_{\bk}} \over{\omega_{\nu{\bk}}}}
\left[e^{-i\omega_{\nu{\bk}}t}(1+n_{\bk})+
e^{i\omega_{\nu{\bk}}t}n_{\bk}\right],\nonumber
\end{eqnarray}
>From (\ref{SS1}) and (\ref{fgexp}) one can easily find explicitly spin-spin
correlator at large separation ${\br}=(m,n)$, $r=\sqrt{m^2+n^2}$
\begin{equation} 
\label{ksi}
\langle {\bf S}(t,{\br}) {\bf S}(t,0) \rangle =
(-1)^{m+n}\exp\left(-{{2\nu}\over{c}}r\right)\times
 \left\{
    \begin{array}{ll}
     1/(\pi c r)^2 ,   & 1 \ll r \ll c\nu /T^2;\\
     2T/(\pi c^3 \nu r), & r \gg c\nu /T^2.
    \end{array} 
  \right.
\end{equation}
So the magnetic correlation length equals $\xi_M=0.5c/\nu$.

Using (\ref{SS1}) and (\ref{fgexp}) we can also find static and dynamic
spin structure factors
\begin{eqnarray}
\label{sffact}
S_M({\bf q})&=&\sum_{\bf r}e^{i{\bf q}\cdot {\bf r}}
\langle {\bf S}(t,{\bf r})\cdot {\bf S}(t,0)\rangle=\\
&=&{2\over {N}}\sum_{\bk}\left[-{1\over{4}}+
{{A_{\bk}A_{\bf k+q}-\gamma_{\bk}\gamma_{\bf k+q}}
\over{\omega_{\nu{\bk}}\omega_{\nu{\bf k+q}}}}
(1+2n_{\bk})(1+2n_{\bf k+q})
\right],\nonumber
\end{eqnarray}
\begin{eqnarray}
\label{dffact}
S_M(\omega,{\bq})&=&\int{{dt}\over{2\pi}}e^{i\omega t}
\sum_{\bf r}e^{i{\bf q}\cdot {\bf r}}
\langle {\bf S}(t,{\bf r})\cdot {\bf S}(0,0)\rangle=\\
&=&{2\over {N}}\sum_{\bk}
\left\{ \left[
-{1\over{4}}+
{{A_{\bk}A_{\bf k+q}-\gamma_{\bk}\gamma_{\bf k+q}}
\over{\omega_{\nu{\bk}}\omega_{\nu{\bf k+q}}}}\right]
\right.
\times \nonumber\\
&\times&
\left[(1+n_{\bk})(1+n_{\bf k+q})
\delta(\omega-\omega_{\nu{\bk}}-\omega_{\nu{\bf k+q}})+
n_{\bk} n_{\bf k+q}
\delta(\omega+\omega_{\nu{\bk}}+\omega_{\nu{\bf k+q}})\right]+\nonumber\\
&+&\left.
\left[{1\over{4}}+
{{A_{\bk}A_{\bf k+q}-\gamma_{\bk}\gamma_{\bf k+q}}
\over{\omega_{\nu{\bk}}\omega_{\nu{\bf k+q}}}}\right]
2 n_{\bk}(1+n_{\bf k+q})
\delta(\omega+\omega_{\nu{\bk}}-\omega_{\nu{\bf k+q}})
\right\}.\nonumber
\end{eqnarray}

We would like to stress that the dynamic structure factor
$S_M(\omega,{\bq})$ at $T=0$ contains only two-quasiparticle
intermediate states. It is similar to the situation in one dimensional
Heisenberg chain where elementary excitation is spinon.  We will
discuss this point in conclusion. Now let us calculate the structure
factors $S_M({\bq})$ and $S_M({\omega,{\bq}})$ for 
${\bq}=0$ and ${\bq}={\bf Q}\equiv (\pm \pi, \pm \pi)$.
Taking into account 
Eq.(\ref{onu}) we find from (\ref{dffact})
\begin{equation}
\label{s0}
S_M(\omega,0)=\delta(\omega) {2\over{N}}\sum_{\bk}n_{\bk}(1+n_{\bk})=
{1\over{\pi c^2}} \delta(\omega) \times
\left\{
\begin{array}{ll}
    \nu_0 T \exp({-\nu_0/T}), & T \ll \nu_0;\\
    1.04 T^2,                     & 1\gg T \gg \nu_0.
\end{array}
\right.
\end{equation}
One can easily obtain static factor $S_M(0)$ integrating (\ref{s0})
by $\omega$. The calculation at ${\bq}= {\bf Q}$ is also very simple.
It gives
\begin{eqnarray}
\label{SQ}
S_M(\omega,{\bf Q})&\approx& {2\over{\pi c^2|\omega|}}
{{1}\over{[1-\exp(-\omega /2T]^2}}
\left[\theta(\omega-2\nu)+\theta(-\omega-2\nu)\right]
+{1\over{2\pi c^2}}\delta(\omega) F\left({{\nu}\over{T}}\right),\\
S_M({\bf Q})& \approx& {2\over{\pi c^2}}\ln{1\over{\nu}},\nonumber
\end{eqnarray}
with function $F$ defined as
\begin{equation}
\label{F}
F\left({{\nu}\over{T}}\right)=\int_{\nu/T}^{\infty}
{{dx}\over{x \sinh^2(x/2)}}=
\left\{
\begin{array}{ll}
    4 T/\nu_0 \exp({-\nu_0/T}), & T \ll \nu_0;\\
    1.67,                     & 1\gg T \gg \nu_0.
\end{array}
\right.
\end{equation}
The $\omega$-distribution at $q=0$ and ${\bq}={\bf Q}$ contains
infinitely sharp $\delta(\omega)$-function and step function 
$\theta(\pm \omega -2\nu)$. However at ${\bq} \ne 0,{\bf Q}$
the $\omega$-distribution is smooth.

Average energy $E=\langle H \rangle$ can be easily calculated
using Eqs. (\ref{H}) and (\ref{fgexp})
\begin{equation}
\label{E}
E/N=2[-g^2(0,{\br}_1)+\alpha f^2(0,{\br}_2)],
\end{equation}
where ${\br}_1=(1,0)$ and ${\br}_2=(1,1)$. As an example, at Fig.1,
we present the plot of energy as a function of temperature at
$\alpha \approx \alpha_c \approx 0.4$. The value of $\alpha-\alpha_c$
is chosen in such a way that the gap at zero temperature equals
$\nu_0=0.05$. In this case our calculation gives for energy per site at 
zero temperature the value $E(T=0)/N=-0.513$. This is slightly above the 
value $E/N \approx -0.520$ found in Ref.\cite{Sch94} by finite lattice
dioganalization and in Ref.\cite{Oit} by series expansion.\\

{\bf Discussion}\\
In the present work using modified spin-wave theory we calculated static 
and dynamic spin structure factors in spin-liquid state of $J_1$-$J_2$ model.
We believe that similar technique can be applied to the $t-J$ model
describing copper oxide superconductors.
Let us point out the strong and weak sides of the modified spin-wave
theory. It gives a very simple description of the spin liquid state, and
this is definitely a strong side. However the description explicitly
violates the rotational symmetry, and this is drawback of the approach.
We would like to stress that violation of exact symmetry quite often
appears in approximate description of a strongly interacting system. For 
example Hartree-Fock method in atoms violates gauge-invariance 
of the electromagnetic transition amplitudes, and
unrestricted Hartree-Fock method in atoms and nuclei violates
rotational symmetry. Usually the symmetry is approximately restored in the
final answer despite the violation on the way. We hope that the
situation is similar in Takahashi's modified spin-wave theory.

In our approach the dynamic structure factor $S_M(\omega,{\bq})$ contains 
only  two-quasiparticle intermediate states. It means that
there is no simple vector excitation which would give 
$C_{\bq}\delta(\omega-\Omega_{\bq})$ contribution to the dynamic structure 
factor. In this sence the picture is similar to that in one dimensional
Heisenberg chain where elementary excitation is spinon.
We stress that the number of dynamic degrees of freedom is also
similar to the $S=1/2$ case because only doublets $S_z=\pm 1$ are 
involved in the  dynamics. However it would be probably wrong to say that 
the spin of elementary excitation $S=1/2$, because still $S_z=\pm 1$. 
More close analogy to the present situation is a gauge vector field in
the axial gauge. In this sence the violation of the rotational
symmetry is similar to the fixing of gauge.
There are two separate questions in this situation\\
1)Is the structure of spectrum in vector channel without narrow
$\delta$-functions valid for $J_1-J_2$ model, or it is
just a byproduct of the approximation? \\
2)Is this technique applicable to $t-J$ model?\\
The answer to first question is not clear. I hope that further
numerical simulations can elucidate this question.
The answer to second question is more clear. 
In $t-J$ model narrow structures in $\omega$-distribution
are much less important due to the very strong damping of spin-waves.
In this situation the most important point is to avoid double
counting of low-energy spin degrees of freedom. 
Takahashi's modified
spin-wave theory definitely provides a correct count of these degrees
of freedom. Therefore in my opinion  this approach is reasonably justified
for $t-J$ model.

\vskip2ex
I would like to thank M. Kuchiev, J. Oitmaa, R. R. P. Singh and
A. Dotsenko for valuable discussions. This work forms part of a project 
supported by a grant of the Australian Research Council.

\tighten

\vskip10ex

Figure caption, Fig.1\\
Energy per site at $J_2/J_1 \equiv \alpha \approx \alpha_c \approx 0.4$. The 
value of $\alpha-\alpha_c$ is chosen in such a way that the gap at zero temperature equals $\nu_0=0.05$.

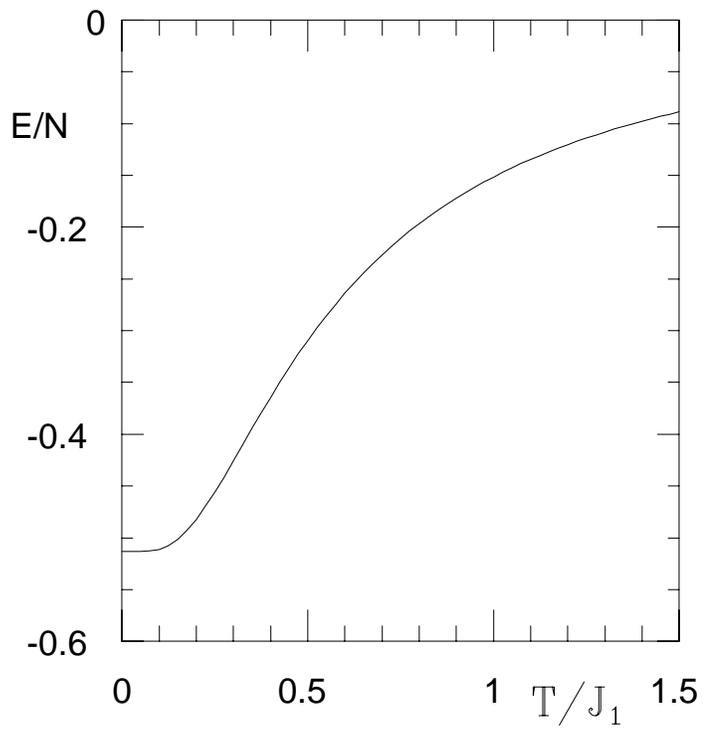
\end{document}